\documentclass[12pt, a4paper]{article}
\usepackage[utf8]{inputenc}
\pdfoutput=1

\usepackage[top=1in, bottom=1in, left=0.75in, right=0.75in]{geometry}
\usepackage[T1]{fontenc}
\usepackage{lmodern}
	\DeclareFontFamily{OMX}{lmex}{}
	\DeclareFontShape{OMX}{lmex}{m}{n}{<-> lmex10}{}	
\usepackage{abstract}
\usepackage{amsmath, amssymb, amsfonts, mathtools}	
	\numberwithin{equation}{section}	
\usepackage{graphicx}       
	\graphicspath{ {./figures/} }
\usepackage[caption=false]{subfig} 
\usepackage[dvipsnames]{xcolor}         
\usepackage{physics}
\usepackage{tensor}
\usepackage{mathrsfs}
\usepackage{bm}
\usepackage{enumitem}
\usepackage{cite}
\usepackage{bigints}    
\usepackage[bottom]{footmisc}	

\usepackage[subfigure]{tocloft}

\usepackage[normalem]{ulem}

\usepackage{hyperref}		
\definecolor{myRed}{rgb}{0.545,0,0}
\definecolor{myDarkBlue}{rgb}{0,0,0.5}
\hypersetup{colorlinks=true, linktocpage, linkcolor=blue, citecolor={myRed}, urlcolor=blue}  

\def \be {\begin{equation}}
\def \ee {\end{equation}}
\def \nn {\nonumber}
\def \del {\partial}

\DeclarePairedDelimiterX\inp[2]{\langle}{\rangle}{#1\,\delimsize\vert\,\mathopen{}#2}	

\begin{document}

\begin{titlepage}
\vspace*{-20mm}   
\baselineskip 10pt    
\begin{flushright}   
\begin{tabular}{r} 
RIKEN-iTHEMS-Report-26
\end{tabular}   
\end{flushright}  
\baselineskip 20pt   
\vglue 10mm

\begin{center}
\noindent
{\LARGE\bf
Exponentially Long Evaporation \\ of Noncommutative Black Hole
\par}
\vskip 10mm
\baselineskip 20pt

\renewcommand{\thefootnote}{\fnsymbol{footnote}}

{\large
Pei-Ming~Ho$^{a, b}$\,\footnote[1]{\texttt{\url{pmho@phys.ntu.edu.tw}}},
Wei-Hsiang~Shao$^{a, c}$\,\footnote[2]{\texttt{\url{whsshao@gmail.com}}},
Takuya~Yoda$^d$\,\footnote[3]{\texttt{\url{t.yoda@gauge.scphys.kyoto-u.ac.jp}}}
}

\renewcommand{\thefootnote}{\arabic{footnote}}

\vskip 5mm

{\small \it  
$^a$Department of Physics and Center for Theoretical Physics, National Taiwan University, \\
No. 1, Sec. 4, Roosevelt Road, Taipei 106319, Taiwan
\\
\vspace{.2cm}
$^b$Physics Division, National Center for Theoretical Sciences, \\
No. 1, Sec. 4, Roosevelt Road, Taipei 106319, Taiwan
\\
\vspace{.2cm}
$^c$RIKEN Center for Interdisciplinary Theoretical and Mathematical Sciences (iTHEMS),\\
RIKEN, Wako 351-0198, Japan
\\
\vspace{.2cm}
$^d$Department of Physics, Kyoto University, 
Kyoto 606-8502, Japan
}  

\vskip 15mm
\begin{abstract}
\vspace{-3mm}
\normalsize

We investigate Hawking radiation in noncommutative spacetime.
For a dynamical black hole formed by the collapse of a matter shell, 
we demonstrate that spacetime noncommutativity modifies the interaction between the radiation field and the background geometry.
In particular, the collapsing shell is effectively shifted by an amount 
proportional to the momentum of an outgoing Hawking mode. 
While the nonlocality inherent in noncommutative spacetime invalidates the conventional arguments for the robustness of Hawking radiation, the radiation decays substantially after the scrambling time, resulting in an exponentially long evaporation time.

\end{abstract}
\end{center}

\end{titlepage}

\newcommand\afterTocSpace{\bigskip\medskip}
\newcommand\afterTocRuleSpace{\bigskip\bigskip}

\hrule
\tableofcontents
\afterTocSpace
\hrule
\afterTocRuleSpace

\newpage 
\section{Introduction}
\label{sec:intro}

Hawking radiation has been a central subject of investigation in theoretical physics since it was predicted in Hawking's seminal works~\cite{Hawking:1974rv, Hawking:1975vcx}. 
Among its profound implications, the black-hole information paradox~\cite{Hawking:1976ra, Hawking:1982dj,Mathur:2009hf,Polchinski:2016hrw,Akil:2025ljb} stands out as one of the most challenging conceptual puzzles. 
In most of the literature addressing this issue, Hawking’s semiclassical analysis~\cite{Hawking:1974rv, Hawking:1975vcx} of particle creation within the framework of low-energy effective theory is regarded robust, provided the black hole remains macroscopically large, with a Schwarzschild radius $a \equiv 2 G_{\text{N}} M$ much larger than the Planck or string length $\ell$.\footnote{In this work, we will not distinguish the Planck scale from the string scale.}
Under this assumption, a black hole is expected to evaporate (almost) completely within a time of the order of magnitude $\mathcal{O}(a^3/\ell^2)$.

Although various general arguments~\cite{Unruh:1977ga, Lowe:1995ac, Hambli:1995pp, Agullo:2006um, Booth:2018xvb} suggest that UV physics should be irrelevant to the Hawking process, it was pointed out in refs.~\cite{Ho:2021sbi, Ho:2024tby} that, with Hawking radiation viewed as a scattering process between outgoing vacuum fluctuations and infalling particles constituting the collapsing matter, the associated center-of-mass energy exceeds the Planck scale after the scrambling time 
$\sim \mathcal{O}\!\left(a\log(a^2/\ell^2)\right)$ 
(see also p.~38 of ref.~\cite{Harlow:2014yka}).\footnote{
Although we refer to the timescale $a \log (a^2 / \ell^2)$ as the “scrambling time,” in the present context this term denotes the critical epoch at which the Hawking process begins to acquire a trans-Planckian character.
Its original definition, which refers to the scrambling of information among black hole degrees of freedom~\cite{Hayden:2007cs, Sekino:2008he}, will not be relevant to our discussion.
}
This locally Lorentz-invariant quantity cannot be dismissed as a coordinate artifact.
Consequently, beyond the scrambling time, Hawking radiation can no longer be reliably described by a low-energy effective theory (see ref.~\cite{Ho:2024tby} for a review).

Nevertheless, the involvement of UV physics in the Hawking process does not necessarily imply that it is sensitive to the UV theory. 
Both the temperature and intensity of Hawking radiation have been shown to remain unchanged (up to small corrections of order $\mathcal{O}(\ell^2 / a^2)$) in most hypothetical UV models considered in the literature~\cite{Unruh:1994je, Brout:1995wp, Corley:1996ar, Corley:1997ef, Corley:1997pr, Barrabes:1998iw, Himemoto:1999kd, Jacobson:1999ay, Barrabes:2000fr, Unruh:2004zk, Coutant:2011in, Robertson:2015cba, Kajuri:2018myh, Boos:2019vcz}. 
Only a few counterexamples have been found to yield $\mathcal{O}(1)$ deviations~\cite{Barcelo:2008qe, Barman:2017vqx, Ho:2022gpg, Akhmedov:2023gqf, Chau:2023zxb, Ho:2023tdq, Ho:2024tby}.

The insensitivity of standard Hawking radiation to the details of the UV models can be understood as follows. As long as the initial quantum state in the asymptotic past redshifts to the local Minkowski (Unruh) vacuum~\cite{Unruh:1976db} 
in the region $\ell \ll r-a \ll a$ (\emph{i.e.}, in the near-horizon region but still many Planck lengths away from the horizon), the standard Hawking process follows, since the subsequent evolution is governed only by low-energy physics. 
In most local UV models studied to date, the state outside the collapsing matter indeed remains close to the Unruh vacuum, which explains the robustness of the standard result. However, this robustness may fail in the presence of nonlocality in the UV theory~\cite{Giddings:2006be}, as has been demonstrated explicitly in models with short-distance cutoffs~\cite{Jacobson:1993hn, Helfer:2003va, Ho:2022gpg} and in frameworks incorporating string-inspired uncertainty relations~\cite{Chau:2023zxb, Ho:2023tdq, Ho:2024tby}.

In particular, it has been suggested~\cite{Ho:2023tdq, Ho:2024tby} that a connection exists between nonlocal theories featuring spacetime uncertainty relations 
and a \emph{short-lived} Hawking radiation. 
For instance, refs.~\cite{Ho:2023tdq, Chang:2024scn} have shown that a Lorentz-covariant version $\Delta u \, \Delta v \gtrsim \ell^2$ of the spacetime uncertainty relation (where $u$ and $v$ are Minkowski light-cone coordinates) can be derived from nonlocal interaction vertices containing infinite-derivative operators of the form $e^{\ell^2 \partial^2}$ in string field theory. 
Such exponential suppression of trans-Planckian interactions was argued to lead to a shutdown of Hawking radiation around the scrambling time $\mathcal{O}\!\left(a\log(a^2/\ell^2)\right)$~\cite{Ho:2023tdq, Ho:2024tby}.  
This observation motivates us to investigate Hawking radiation in other models exhibiting a similar spacetime uncertainty relation. 
Since the spacetime uncertainty relation has been proposed as a fundamental property of string theory~\cite{Yoneya:1987gb, Yoneya:1989ai, Yoneya:1997gs, Yoneya:2000bt}, this direction of study may shed light on how certain nonlocal features of string theory influence Hawking radiation.

A natural realization of the spacetime uncertainty relation is through noncommutative field theories, which appear in string theory as the low-energy effective dynamics of open strings in the presence of a background $B$-field on D-branes~\cite{Douglas:1997fm, Chu:1998qz, Seiberg:1999vs}. 
Equivalent noncommutative structures also emerge from the algebra of open string field theory~\cite{Witten:1985cc} and from matrix models~\cite{Banks:1996vh, Ishibashi:1996xs, Connes:1997cr}.  
In this work, we employ noncommutative field theory as a toy model for the radiation field to realize the spacetime uncertainty relation and to study its effects on Hawking radiation.

We remark that noncommutative field theories typically contain infinitely many time derivatives in their actions, which may lead to apparent violations of unitarity~\cite{Gomis:2000zz, Chaichian:2000ia, Alvarez-Gaume:2001dfr, Chu:2002fe, Bahns:2002vm} and causality~\cite{Seiberg:2000gc, Chaichian:2000ia, Alvarez-Gaume:2001dfr, Chu:2005nb}.
In this work, however, we implement the noncommutativity in a controlled manner so as to avoid these pathologies, at least for the purpose of analyzing Hawking radiation.\footnote{
Recall that, while string theory is well-defined, the low-energy truncation of string field theory~\cite{Witten:1985cc} also contains infinite time derivatives and may appear to exhibit similar pathologies at first sight~\cite{Eliezer:1989cr}.
} 
Although our calculation is far from a full-fledged derivation from string theory, extracting specific features of string theory (in this case the spacetime uncertainty relation) and embedding them into simplified effective models may provide valuable insight into the behavior of Hawking radiation in quantum gravity.

Most prior studies of noncommutative black holes~\cite{Nicolini:2005vd, Banerjee:2008gc, Nozari:2008rc, Modesto:2010uh} do not really employ a noncommutative spacetime algebra.
Instead, they model noncommutative effects by replacing point sources (such as the central singularity) with smeared distributions. 
Naturally, these constructions lead only to small deviations from the standard Hawking result until the very late stage of evaporation. 
In contrast, the present work considers a setting in which the spacetime coordinates obey a noncommutative algebra.

A priori, it is not obvious how spacetime noncommutativity should be incorporated into black-hole physics, since its implementation depends on the choice of coordinates.
Our approach is motivated by recent works~\cite{Aoude:2024sve, Aoki:2025ihc} (see also refs.~\cite{Ilderton:2025aql, Aoude:2025jvt}) in which Hawking radiation is reformulated in terms of scattering amplitudes.
In this framework, the standard Hawking spectrum is reproduced by analyzing how an outgoing probe particle scatters off the collapse geometry, which is treated perturbatively as an external gravitational field in Minkowski spacetime.
It is then natural to implement noncommutativity (at the Planck scale $\ell^{-1}$) in the underlying Minkowski spacetime of this formulation, and include the gravitational background as a potential term in the noncommutative field theory. 
Crucially, low-energy physics is unchanged for both freely falling and distant observers, whereas the nonlocality associated with the noncommutative geometry becomes macroscopic only at trans-Planckian energies.

Treating these noncommutative effects nonperturbatively in the Planck length $\ell$, we find that the noncommutative model reproduces the standard Hawking temperature, but predicts a time-dependent radiation intensity that becomes strongly suppressed after the scrambling time $\mathcal{O}(a \log(a^2/\ell^2))$. In contrast to previous studies of nonlocal UV physics~\cite{Ho:2022gpg, Chau:2023zxb, Ho:2023tdq}, in which Hawking radiation terminates at around the scrambling time (see ref.~\cite{Ho:2024tby} for a review), the noncommutative black hole in our model continues to evaporate over a finite yet exponentially long timescale $\sim \exp(a^2/\ell^2)$.

This paper is organized as follows. 
In section~\ref{sec:HR0}, we introduce the basic setup and review the derivation of Hawking radiation from a black hole formed by the dynamical collapse of a matter shell, following Hawking’s original analysis~\cite{Hawking:1974rv,Hawking:1975vcx} in the low-energy effective theory. 
In section~\ref{sec:NC}, the quantum theory of the radiation field is extended to a noncommutative background as a toy model for the UV physics. 
In this framework, the background geometry effectively depends on the momentum of the propagating mode.
We then compute the resulting Hawking radiation in section~\ref{sec:HR_NC} and show that its intensity is significantly suppressed after the scrambling time, while the total evaporation time is of the order of the Poincar\'e recurrence time, far exceeding the Page time $\mathcal{O}(a^3/\ell^2)$~\cite{Page:1993wv}.
Finally, we summarize and discuss the implications of our findings in section~\ref{sec:conclude}.
Throughout the paper we adopt natural units $c = 1$, $\hbar = 1$, and $k_{\mathrm{B}} = 1$, where $k_{\mathrm{B}}$ is the Boltzmann constant.

\section{Hawking Radiation in Low-Energy Effective Theory}
\label{sec:HR0}

In this section, we review the derivation of Hawking radiation within the framework of low-energy effective field theory.
(One may refer to ref.~\cite{Ho:2024tby} for more details.)

Consider a massless scalar field $\varphi$ propagating on an ingoing Vaidya background~\cite{Griffiths:2009dfa}:
\be
S_0 [\varphi] 
= 
-\frac{1}{2} 
\int d^4 x \sqrt{-g} \,
g^{\mu\nu} 
\left( \partial_\mu \, \varphi \right) 
\left( \partial_\nu \, \varphi \right) 
,
\label{S0_varphi}
\ee
where the Vaidya metric is given by
\be
ds^2
= 
-\left( 1-\frac{a\Theta(v)}{r} \right) 
dv^2
+ 2 \, dv \, dr
+ r^2 d \Omega^2
\, .
\label{vaidya}
\ee
Here, $\Theta(v)$ denotes the step function, and $a$ is the Schwarzschild radius defined by 
\be
a=2 G_{\text{N}} M
\, ,
\ee
with $G_{\text{N}}$ being the Newton constant and $M$ the mass of the collapsing matter.
For simplicity, we model the collapse by a spherically symmetric lightlike thin shell located at $v = 0$, where $v$ is the ingoing Eddington-Finkelstein coordinate.
The spacetime is Minkowski for $v < 0$ and Schwarzschild for $v > 0$.

The coordinates $(v, r)$ are convenient because they coincide with the Minkowskian advanced time $v$ and radial coordinate $r$ both inside the collapsing shell and in the asymptotically flat region ($r \rightarrow \infty$) outside the shell.
The metric~\eqref{vaidya} can be decomposed as the Minkowski metric 
\be
\label{ds^2_flat}
ds_{\text{flat}}^2 = - dv^2 + 2 \, dv \, dr
+ r^2 d \Omega^2
\ee
plus a correction term $B(v, r) \, dv^2$ from the gravitational background,
with
\be
B(v, r) \equiv \frac{a\Theta(v)}{r} \, .
\label{B(v, r)}
\ee

To simplify the analysis, we restrict our attention to $s$-wave configurations $\varphi = \varphi(v, r)$, and introduce a two-dimensional scalar field $\phi(v, r)$ via
\be
\varphi(v, r) = \frac{\phi(v, r)}{\sqrt{4\pi} r}
\, ,
\ee
in terms of which the action~\eqref{S0_varphi} is brought to the form 
\be 
\label{S0_phi}
S_0 [\phi(v, r)]
= 
- \frac{1}{2} 
\int dv \int dr \, 
\Bigl\{
2 \left( \del_v \, \phi \right)
\left( \del_r \, \phi \right) 
+ 
\left[ 1 - B(v, r) \right]
(\del_r \, \phi)^2 
+
V(v, r) \, \phi^2
\Bigr\}
\, .
\ee 
where $V(v, r) = B(v, r) / r^2$ acts as a potential barrier.

The potential barrier $V(v,r)$ induces greybody factors~\cite{Hawking:1975vcx, Page:1976df} and thus slightly distorts the Hawking spectrum from an exact Planck distribution. 
It can be neglected for an order-of-magnitude estimate, 
as in Hawking’s original derivation~\cite{Hawking:1975vcx}. 
Since $V(v,r)$ contributes only subleading corrections 
of order $\mathcal{O}(\ell^2/a^2)$ (with $\ell$ denoting the Planck length),
this omission is not expected to affect the main conclusions of this work regarding large modifications to Hawking radiation in the noncommutative field theory,
which are $\mathcal{O}(1)$ with respect to the perturbative parameter $\ell / a$ 
(see section~\ref{sec:HR_NC}).

With this simplification, the action~\eqref{S0_phi} reduces to the two-dimensional form
\be
S_{\text{2D}} [\phi]
= 
- \frac{1}{2} 
\int dv \int dr 
\left[
2 \left( \del_v \, \phi \right)
\left( \del_r \, \phi \right) 
+ 
(\del_r \, \phi)^2 
- 
B(v, r) \, (\del_r \, \phi)^2 
\right]
.
\label{S0_phi_2}
\ee
The corresponding equation of motion is
\be
\del_r 
\left[2 \del_v \, \phi 
+ 
\del_r \, \phi 
- 
B(v, r) \, \del_r \, \phi
\right] 
= 
0 \, .
\ee
For the analysis of Hawking radiation, 
it is sufficient to consider the equation governing the outgoing modes,
which reads
\be
\label{wave_eq0}
\left[ 
2\del_v + \del_r - B(v, r) \, \del_r \right] 
\phi = 0
\, .
\ee

\subsection{Solution to Wave Equation}
\label{sec:sol0}

In the near-horizon region $r - a \ll a$ outside the shell $(v > 0)$, 
the single-frequency solution to eq.~\eqref{wave_eq0}
is approximately given in momentum space by~\cite{Akhmedov:2023gqf, Chau:2023zxb, Ho:2024tby}\,\footnote{
The $i\varepsilon$ prescription ensures that the solution~\eqref{phi0_om} 
has support only outside the horizon, 
thereby respecting the boundary conditions 
appropriate for an outgoing Hawking mode.
}
\be 
\label{phi0_om}
\tilde{\phi}_{\omega}(v, p)
\equiv 
\mathcal{N}_{\omega} \, 
e^{-i \omega v} \, 
e^{- iap} 
\left[ 
a \left( p - i \varepsilon \right) 
\right]^{-1 \, - \, 2 i a \omega}
\qquad 
(\varepsilon \to 0^+) \, ,
\ee 
where $\tilde{\phi}(v,p)$ denotes the Fourier transform of the field:
\be
\phi(v, r) = \int_{-\infty}^{\infty} \frac{dp}{\sqrt{2 \pi}} \, \tilde{\phi} (v, p) \, e^{ipr} \, ,
\label{Fourier}
\ee
and the normalization factor is
\be 
\label{N}
\mathcal{N}_{\omega}
=
\frac{a^2}{\pi} 
\sqrt{\frac{\omega}{2}} \, e^{\pi a \omega} \, \Gamma(2 i a \omega) \, .
\ee

Inside the shell ($v < 0$), 
the quantum field obeying the outgoing wave equation
\be
\label{wave_eq0_inn}
\left(2\del_v + \del_r\right) \phi = 0
\ee
admits the mode expansion
\be 
\label{phi0_in}
\Tilde{\phi} (v, p)
=
\frac{e^{-i p v / 2}}{\sqrt{2 \abs{p}}} \, 
\bigl[ \mathfrak{a}_p \, \Theta(p) + \mathfrak{a}_{-p}^{\dagger} \, \Theta(-p) \bigr] 
\qquad 
(v < 0) \, ,
\ee 
where $\mathfrak{a}_p$ and $\mathfrak{a}_p^{\dagger}$ are the annihilation and creation operators satisfying the canonical commutation relations
\be 
\label{a_comm}
[ \mathfrak{a}_p \, , \mathfrak{a}_{p'}^{\dagger} ]
=
\delta(p - p')
\, ,
\qquad 
\left[ \mathfrak{a}_p \, , \mathfrak{a}_{p'} \right]
=
0
\, , 
\qquad 
[ \mathfrak{a}_p^{\dagger} \, , \mathfrak{a}_{p'}^{\dagger} ]
=
0
\, .
\ee 
The initial state is assumed to be the Minkowski vacuum $\ket{0}$ inside the shell, 
defined by
\be
\label{vacuum}
\mathfrak{a}_p \ket{0} = 0 \qquad \forall \ p > 0 \, .
\ee

To analyze the time dependence of Hawking radiation, it is necessary to work with localized wave packets~\cite{Hawking:1975vcx} rather than 
the single-frequency modes $\tilde{\phi}_{\omega}$~\eqref{phi0_om},
which occupy the entire spacetime region outside the horizon. 
A generic wave packet describing the field configuration outside the shell 
can be written as
\be 
\label{packet0}
\tilde{\phi} (v > 0, p)
=
\widetilde{\Psi}_{\omega_c , \, u_c} (v, p)
\equiv 
\int_{-\infty}^{\infty} 
d \omega \, 
\widetilde{\psi}_{\omega_c}(\omega) \, 
e^{i \omega u_c} \, 
\tilde{\phi}_{\omega}(v, p) \, ,
\ee 
namely, as a superposition of 
the single-frequency solutions $\tilde{\phi}_{\omega}(v,p)$~\eqref{phi0_om}
weighted by the profile function $\widetilde{\psi}_{\omega_c}(\omega)$ 
in frequency space.
The profile $\widetilde{\psi}_{\omega_c}(\omega)$ 
is assumed to have compact support 
around a positive central frequency $\omega=\omega_c>0$, 
with a narrow width $\Delta\omega\ll\omega_c$. 
The wave packet $\widetilde{\Psi}_{\omega_c,u_c}$ is therefore 
composed of purely positive-frequency modes with respect to a distant observer, 
and we use it to represent the wavefunction of a particle 
carrying asymptotic frequency $\omega_c$ in the Hawking radiation 
detected by such an observer.

It can be shown that the wave packet~\eqref{packet0} has the asymptotic form~\cite{Chau:2023zxb}
\be 
\Psi_{\omega_c , \, u_c} (v, r \to \infty)
\propto 
\int_{-\infty}^{\infty} 
d \omega \, 
\widetilde{\psi}_{\omega_c}(\omega) \, 
e^{- i \omega \, \left[ u(v, \, r) \, - \, u_c \right]} 
\ee 
in position space,
where 
\be 
\label{u(v,r)}
u(v,r)
=
v-2r_*
\equiv 
v-2r-2a \log\biggl(\frac{r}{a}-1\biggr)
\ee 
denotes the Eddington retarded time.
It is now clear that the phase factor $e^{i \omega u_c}$ introduced in eq.~\eqref{packet0} plays the role of shifting the central retarded time location of the wave packet in the asymptotic region from $u = 0$ to $u = u_c$. 
It is the introduction of this central time parameter $u_c$ that will allow us to capture the nontrivial time dependence of Hawking radiation in the presence of UV effects, as has been demonstrated in a series of past works~\cite{Jacobson:1993hn, Helfer:2003va, Barcelo:2008qe, Barman:2017vqx, Ho:2022gpg, Akhmedov:2023gqf, Chau:2023zxb, Ho:2023tdq, Ho:2024tby}. 

The annihilation operator $\mathfrak{b}_{\Psi}$ associated with a given particle wavefunction~\eqref{packet0} can be defined through the relativistic inner product
\be
\langle \Psi_1 \, , \Psi_2 \rangle 
\equiv 
- i \int_{-\infty}^{\infty} dr \, 
\Psi_1^{\ast}(v, r) \, \overset{\leftrightarrow}{\del_r} \, \Psi_2 (v, r)
=
\int_{-\infty}^{\infty} dp \, 
\widetilde{\Psi}_1^{\ast} (v, p) \, (2p) \, \widetilde{\Psi}_2 (v, p)
\label{inner}
\ee
as
\be 
\label{b_Psi}
\mathfrak{b}_{\Psi} 
\equiv 
\bigl\langle 
\Psi_{\omega_c, \, u_c} \, , 
\phi
\bigr\rangle 
\, .
\ee
With the frequency profile $\widetilde{\psi}_{\omega_c} (\omega)$ normalized according to 
\be
\int_{-\infty}^{\infty} d\omega \, 
\bigl| \widetilde{\psi}_{\omega_c} (\omega) \bigr|^2
= 
1 
\, ,
\ee
the operators $\mathfrak{b}_{\Psi}$ and $\mathfrak{b}_{\Psi}^{\dagger}$ obey the canonical commutation relation $[ \mathfrak{b}_{\psi} \, , \mathfrak{b}^{\dagger}_{\psi} ] = 1$. 

\subsection{Hawking Particle Number}
\label{sec:vev0}

We can now evaluate the physical observable relevant for Hawking radiation, 
namely the vacuum expectation value 
$\ev{\mathfrak{b}_{\Psi}^{\dagger} \mathfrak{b}_{\Psi}}{0}$
of the number operator $\mathfrak{b}_{\Psi}^{\dagger} \mathfrak{b}_{\Psi}$
associated with a generic wave packet $\widetilde{\Psi}_{\omega_c,u_c}$~\eqref{packet0}.

By imposing continuity of the field across the shell at $v=0$ 
and using the wave equation~\eqref{wave_eq0_inn} inside the shell,
one finds that a Hawking wave packet~\eqref{packet0} takes the form
\be 
\label{packet0_I}
\widetilde{\Psi}_{\omega_c, \, u_c} ( v < 0 \, , p)
=
e^{-i p v / 2} \;
\widetilde{\Psi}_{\omega_c, \, u_c} ( 0 \, , p)
\ee 
when traced back in time to the interior of the shell. 
Taking its inner product~\eqref{b_Psi} 
with the mode expansion~\eqref{phi0_in} inside the shell 
then yields the Bogoliubov transformation
\be 
\mathfrak{b}_{\Psi} 
=
\int_0^{\infty} dp \, 
\sqrt{2 p} \,
\bigl[
e^{-i p v / 2} \; 
\widetilde{\Psi}_{\omega_c, \, u_c}^{\ast} (0 \, , p ) \, 
\mathfrak{a}_p 
- 
e^{i p v / 2} \; 
\widetilde{\Psi}_{\omega_c, \, u_c}^{\ast} (0 \, , - p ) \, 
\mathfrak{a}_p^{\dagger} 
\bigr]
\, .
\ee 
As a result, the expectation value of the number operator is
\be 
\label{bbvev0}
\ev{\mathfrak{b}_{\Psi}^{\dagger} \mathfrak{b}_{\Psi}}{0} 
=
\int_0^{\infty} dp 
\left( 2 p \right) 
\bigl| 
\widetilde{\Psi}_{\omega_c, \, u_c} (0 \, , - p)
\bigr|^2
\, ,
\ee 
which is essentially computing the norm of the negative-momentum components of the wave packet $\widetilde{\Psi}_{\omega_c, \, u_c}$ when propagated back to the shell surface at $v = 0$. 

To evaluate the integral in eq.~\eqref{bbvev0}, further approximations shall be made based on the following observation: Given a wave packet $\widetilde{\Psi}_{\omega_c, \, u_c}$~\eqref{packet0} with a sufficiently narrow frequency profile $\widetilde{\psi}_{\omega_c} (\omega)$, we can pull out factors in the integrand of~\eqref{packet0} that varies slowly with $\omega$ within a small neighborhood $\Delta \omega \ll \omega_c$ around the central frequency $\omega = \omega_c$ of the packet.
This allows us to approximate the negative-$p$ components of the wave packet as\,\footnote{
In general, given a wave packet profile $\widetilde{\psi}_{\omega_c}(\omega)$ with width $\Delta \omega$ in the frequency space, the following approximation
\be 
\int_{-\infty}^{\infty} d \omega \, 
\widetilde{\psi}_{\omega_c}(\omega) 
\prod_{i = 1}^N 
f_i (\omega)
\simeq 
\prod_{i = 1}^N 
f_i (\omega_c)
\int_{-\infty}^{\infty}
\widetilde{\psi}_{\omega_c}(\omega) \,
d \omega 
\nn 
\ee 
is valid if $\abs{f_i^{\prime} (\omega_c) / f_i (\omega_c)} \, \Delta \omega \ll 1$ for all $i \in \{ 1, 2, \cdots , N \}$.
In the case of eq.~\eqref{packet_approx}, it can be shown that the condition under which the approximation scheme is justified is $\Delta \omega \ll a^{-1} \sim \omega_c$.
}
\begin{align}
\widetilde{\Psi}_{\omega_c, \, u_c} (v, - \abs{p})
&=
\int_{-\infty}^{\infty} d \omega \, 
\widetilde{\psi}_{\omega_c}(\omega) \, 
e^{- i \omega \left( v \, - \, u_c \right)} \, 
\tilde{\phi}_{\omega}(- \abs{p})
\nn \\
&\simeq
-
\mathcal{N}_{\omega_c} \, 
e^{-2 \pi a \, \omega_c} \;
\frac{e^{i a \abs{p}}}{a \abs{p}} 
\int_{-\infty}^{\infty} d \omega \, 
\widetilde{\psi}_{\omega_c}(\omega) \, 
e^{- i \omega \, \left[ v \, - \, u_c \, + \, 2 a \log( a \abs{p} ) \right]} 
\, .
\label{packet_approx}
\end{align}

With this in mind, and recalling from eq.~\eqref{N} that 
\be 
\abs{\mathcal{N}_{\omega}}^2
=
\frac{a^3}{2 \pi} \, 
\frac{1}{1 - e^{- 4 \pi a \, \omega}}
\, ,
\ee 
the number expectation value~\eqref{bbvev0} 
can be further expressed as 
\be 
\label{bbvev0_1}
\ev{\mathfrak{b}_{\Psi}^{\dagger} \mathfrak{b}_{\Psi}}{0} 
\simeq 
\frac{2 a}{e^{4 \pi a \, \omega_c} - 1}
\int_0^{\infty} 
\frac{dp}{p} \, 
\abs{\psi_{\omega_c} 
\bigl( u_c - 2 a \log (a p) \bigr)}^2 \, ,
\ee 
where we have defined 
\be 
\label{psi(u)}
\psi_{\omega_c} (u)
\equiv 
\int_{-\infty}^{\infty} 
\frac{d \omega}{\sqrt{2 \pi}} \, 
\widetilde{\psi}_{\omega_c} (\omega) \, 
e^{i \omega u}
\ee 
to be the Fourier counterpart of the frequency profile 
$\widetilde{\psi}_{\omega_c} (\omega)$.
Written in this form, the profile $\psi_{\omega_c} (u)$ 
is centered around $u = 0$.
Finally, by performing the change of variable
\be 
\label{p_to_u(p)}
p \ \mapsto \ 
u_0 (p; u_c) 
\equiv 
u_c - 
2 a \log(ap)
\ee 
in eq.~\eqref{bbvev0_1}, we arrive at
\be 
\label{bb_final0}
\ev{\mathfrak{b}_{\Psi}^{\dagger} \mathfrak{b}_{\Psi}}{0} 
\simeq 
\frac{1}{e^{4 \pi a \, \omega_c} - 1}
\int_{-\infty}^{\infty} 
du_0 \, 
\abs{\psi_{\omega_c} (u_0)}^2
=
\frac{1}{e^{4 \pi a \, \omega_c} - 1}
\, ,
\ee 
which reflects a time-independent (\emph{i.e.}, $u_c$-independent) magnitude of Hawking radiation at the temperature $T_{\mathrm{H}} \simeq (4 \pi a)^{-1}$. 
This completes the derivation of the standard Hawking radiation in the low-energy effective theory.

\newpage 
\section{Noncommutative Black-Hole Geometry}
\label{sec:NC}

In this section, we investigate the effects of spacetime noncommutativity on Hawking radiation. 
On the Vaidya geometry~\eqref{vaidya}, 
we introduce noncommutativity through the Moyal star product~\cite{Groenewold:1946kp, Moyal:1949sk}:
\be
\label{star}
(f\star g)(v, r)
\equiv
\left.
\exp\left[
\frac{i\ell^2}{2} 
\left( \del_v \, \del_{r'} - \del_r \, \del_{v'} \right)
\right]
f(v, r) \, g(v', r')
\right|_{v' \, = \, v , \, 
r' \, = \, r}
\, .
\ee 
The noncommutativity defined in this way 
gives rise to the spacetime uncertainty relation~\cite{Szabo:2001kg}
\be
\label{stur_vr}
\Delta v \, 
\Delta r \gtrsim \ell^2
\, .
\ee
Expressed in terms of the Eddington lightcone coordinates $(u, v)$, 
this implies the uncertainty relation $\Delta u \, \Delta v \gtrsim 2 \ell^2$
in the asymptotically flat region.
Likewise, near the horizon, one finds
\be
\Delta U \, \Delta V \gtrsim 2 \ell^2
\label{stur_UV}
\ee
in terms of the Kruskal coordinates $U(u) \equiv - 2a \, e^{-u / 2a}$ and $V(v) \equiv 2a \, e^{v / 2a}$.\footnote{
The spacetime uncertainty relation~\eqref{stur_UV} 
coincides in form with that proposed by Yoneya~\cite{Yoneya:1987gb, Yoneya:1989ai, Yoneya:1997gs, Yoneya:2000bt} 
as a fundamental principle of string theory, 
and was also derived recently in refs.~\cite{Ho:2023tdq, Chang:2024scn} 
from the nonlocal structure of string field theory.
}

For both distant and freely falling observers, 
such effects can only be detected in Planck-scale experiments. 
Interestingly, there can be macroscopic nonlocality $\Delta V \gg a$ 
in the trans-Planckian (short-distance) regime where $\Delta U \ll \ell^2/a$.
Consequently, spacetime noncommutativity can significantly modify 
the evolution of outgoing modes with trans-Planckian momenta, 
and hence the late-time behavior of Hawking radiation.

Implementing noncommutativity~\eqref{star} in a different coordinate system 
may lead to a distinct theory.
Nevertheless, we expect the conclusion of this work 
regarding the late-time modification of Hawking radiation
to remain qualitatively the same as long as 
spacetime noncommutativity is present at a finite scale $\ell$ near the horizon.
Furthermore, 
although spacetime noncommutativity typically leads to issues with causality, 
we will incorporate the star product~\eqref{star} into the wave equation for $\phi$ 
in a manner consistent with causality.

\subsection{Noncommutative Wave Equation}

The wave equation~\eqref{wave_eq0} for outgoing modes 
in the commutative Vaidya spacetime 
with a collapsing null shell at $v=0$ is
\be
\label{wave_eq0_rewrite}
(2\partial_v+\partial_r) \, \phi 
- 
B(v, r) \, \del_r \, \phi = 0 
\, .
\ee 
In particular, the term $B(v,r)\,\partial_r \, \phi$ 
may be interpreted as an effective gravitational potential 
produced by the collapsing matter in a flat background. 
Indeed, Hawking radiation can in principle be derived 
in the framework of scattering amplitudes 
by treating the term $B(v, r) \, \del_r \, \phi$ perturbatively 
as a background interaction in flat spacetime~\eqref{ds^2_flat}~\cite{Aoude:2024sve}. 
This perspective naturally motivates our implementation of noncommutativity~\eqref{star} 
in the dynamical black-hole geometry.

A natural noncommutative deformation of the wave equation~\eqref{wave_eq0_rewrite} is
\be
(2\partial_v+\partial_r) \, 
\phi
-
\frac{1}{2}
\left[
B(v, r) \star \partial_r \, \phi
+
\partial_r \, \phi \star B(v, r)
\right] 
= 
0
\, ,
\label{eom_1}
\ee 
where the interaction term has been symmetrized 
in order to preserve the reality of the scalar field $\phi$.
After Fourier transformation~\eqref{Fourier}, 
this equation is equivalent to
\be 
(2 \del_v + ip) \, \tilde{\phi} (v, p) 
= 
\frac{ia}{2} \, 
\Biggl[
\Theta\biggl(v - \frac{\ell^2 p}{2} \biggr) \, 
\frac{1}{r - i \ell^2 \del_v / 2} 
+ 
\Theta\biggl(v + \frac{\ell^2 p}{2} \biggr) \, 
\frac{1}{r + i \ell^2 \del_v / 2} 
\Biggr] 
\, p \, \tilde{\phi} (v, p) \, ,
\label{eom_1p}
\ee 
where we have used the identity 
\begin{align}
\frac{a\Theta(v)}{r} \star \left[ ip \, e^{i p r} \, \tilde{\phi} (v, p) \right]
&=
iap \left\{e^{i p r'} \left[ e^{- \ell^2 p \, \del_v / 2} \, \Theta(v) \right] e^{- i \ell^2 \del_r \, \del_{v'}} \, \frac{\tilde{\phi}(v', p)}{r}\right\}_{v' \, = \, v, \, r' \, = \, r}
\nn 
\\
&=
\frac{i a p}{r} \, \Theta\biggl(v - \frac{\ell^2 p}{2} \biggr) \left[\frac{1}{1 - i \ell^2 \del_v / 2r} \, \tilde{\phi}(v, p)\right] e^{ipr} \, ,
\end{align}
as well as the analogous identity
\be 
\left[ ip \, e^{i p r} \, \tilde{\phi} (v, p) \right] \star \frac{a\Theta(v)}{r} 
= \frac{i a p}{r} \, \Theta\biggl(v + \frac{\ell^2 p}{2} \biggr) \left[\frac{1}{1 + i \ell^2 \del_v / 2r} \, \tilde{\phi}(v, p)\right] e^{ipr}
\ee 
for the term $(\del_r \, \phi) \star B(v, r)$.
In the momentum-space representation~\eqref{eom_1p}, 
the radial coordinate $r$ should be understood 
as the derivative operator conjugate to $p$.
It acts not only on $p\,\tilde{\phi}(v,p)$, 
but also on the $p$-dependence of the step functions to its left.

The noncommutative wave equation~\eqref{eom_1p} suffers from two drawbacks.
First, the step functions $\Theta(v - \ell^2 p / 2)$ and $\Theta(v + \ell^2 p / 2)$
suggest that the collapsing thin shell at $v=0$ is effectively split into two thin shells located at $v=\pm \ell^2 p/2$, or equivalently at $v=\pm \ell^2 |p|/2$.
The shell at $v=-\ell^2 |p|/2$ leads to an apparent violation of causality.
To see this, suppose that the collapsing matter is released at $v=0$ 
by experimentalists far from the black hole.
For an outgoing mode with large momentum $p$, 
the wave equation~\eqref{eom_1p} implies the presence of a shell 
already at the earlier time $v=-\ell^2 |p|/2<0$, 
even though the experimentalists could in principle 
still choose to abort the collapse at any time 
between $v=-\ell^2 |p|/2$ and $v=0$. 
The equation is therefore acausal in this sense.

Second, the factors $1/(r \pm i \ell^2 \del_v / 2)$ in eq.~\eqref{eom_1p} 
become singular at large $r$ when the frequency $\omega = i \del_v$ is large.
Although the calculation of Hawking radiation concerns only low-energy modes 
with $\omega\sim\mathcal{O}(1/a)>0$ outside the split shells ($v > \ell^2 |p| / 2$)
where $1/(r \pm i \ell^2 \del_v / 2)$ remain finite, 
the corresponding frequency $\omega$ inside the shell(s) 
is expected to be trans-Planckian due to the large blueshift.
Hence, if the wave equation~\eqref{eom_1} is to be regarded as a sensible UV model
allowing arbitrarily large $\omega$, this large-$r$ singularity must be removed.
We do so below by replacing the star product acting on the factor $1/r$ in the gravitational potential with the ordinary product.

To obtain a causal wave equation on noncommutative spacetime,
we first decompose the scalar field $\phi$ into its 
positive- and negative-momentum components as
\be 
\phi(v, r) = \phi^{(+)}(v, r) + \phi^{(-)}(v, r) \, ,
\qquad
\text{where}
\quad 
\phi^{(\pm)}(v, r) \equiv 
\int_0^{\infty} \frac{dp}{\sqrt{2 \pi}} \, 
\tilde{\phi}^{(\pm)}(v, \pm p) \, 
e^{\pm ipr} \, .
\ee 
The reality condition $\phi^{\ast}(v, r) = \phi(v, r)$ then implies 
\be 
\left[ \tilde{\phi}^{(+)}(v, p)\right]^{\ast} = \tilde{\phi}^{(-)}(v, -p)
\qquad (p > 0) \, .
\ee 
To avoid the singularity and acausality issues discussed above, 
we consider the deformed equation of motion
\be 
(2\partial_v+\partial_r) \, \phi^{(+)}(v, r) 
- 
\frac{a}{r} 
\left[ \Theta(v) \star \partial_r \, \phi^{(+)}(v, r)\right] 
= 0 
\ee 
for outgoing modes with positive momentum.\footnote{
For the analysis of Hawking radiation, 
the qualitative conclusions do not depend on 
whether the factor $1/r$ in the gravitational potential is introduced 
via the ordinary product or the star product.
}
In momentum space, this becomes
\be 
(2\partial_v+ip) \, \tilde{\phi}^{(+)}(v, p) 
- \frac{ia}{r} \, 
\Theta\biggl( v- \frac{\ell^2 p}{2} \biggr) \, 
p \, \tilde{\phi}^{(+)}(v, p) = 0
\qquad (p > 0) 
\, .
\label{eom_2p>0}
\ee 
The equation for negative-momentum modes 
is then fixed by complex conjugation, yielding
\be 
(2\partial_v-ip) \, \tilde{\phi}^{(-)}(v, -p) 
+ 
\frac{ia}{r} \, 
\Theta\biggl( v- \frac{\ell^2 p}{2} \biggr) \, 
p \, \tilde{\phi}^{(-)}(v, -p) = 0
\qquad (p > 0) \, .
\label{eom_2p<0}
\ee

The noncommutative wave equation for both positive- and negative-momentum modes 
can therefore be written as 
\be 
(2\partial_v+ip) \, \tilde{\phi}(v, p) 
- \frac{ia}{r} \, 
\Theta\biggl( v- \frac{\ell^2 |p|}{2} \biggr)\, 
p \, \tilde{\phi}(v, p) 
= 0
\qquad (p \in \mathbb{R}) \, ,
\label{eom_2p}
\ee 
which is manifestly causal.
There is now only a single thin shell, 
whose location is shifted forward in the advanced time $v$ by
\be 
\Delta v
=
\frac{\ell^2 |p|}{2}
\ee 
for an outgoing mode with momentum $p$. 
This shift reflects a UV-IR connection inherent in spacetime noncommutativity~\eqref{stur_vr}, since $\Delta v$ becomes macroscopic 
in the UV limit $|p|\to\infty$.
This momentum-dependent deformation of the effective background geometry is a general feature of theories exhibiting spacetime uncertainty relations, as has been observed in various other contexts~\cite{Brandenberger:2002nq, Chang:2024scn}.

Strictly speaking, since $r$ is defined as a Hermitian operator 
acting on functions of $p$, 
the operator ordering between $r$ and $p$ in the wave equation~\eqref{eom_2p} 
requires some care. 
More precisely, the wave equation~\eqref{eom_2p} 
should be regarded as an equation of the form
\be
i\partial_v \, \tilde{\phi}(v, p) = H \tilde{\phi}(v, p)
\, ,
\label{eom_2p_H}
\ee
where $H$ is a Hamiltonian generating evolution in $v$ 
and must be Hermitian with respect to the relativistic inner product~\eqref{inner}. 
One simple choice of operator ordering is
\be
H \equiv
\frac{p}{2} 
- \frac{a}{2\sqrt{r}} \, 
\Theta\biggl( v- \frac{\ell^2 |p|}{2} \biggr) \, 
\frac{1}{\sqrt{r}} \, p \, .
\label{H}
\ee
As will become clear in the discussion below,
this modification will not significantly alter the prediction of Hawking radiation.
We therefore proceed with the simpler equation~\eqref{eom_2p}.

\subsection{Semiclassical Approximation}
\label{sec:semiclassical}

Before solving the noncommutative wave equation~\eqref{eom_2p} 
and deriving the corresponding Hawking radiation, 
we first analyze the wave equation in the semiclassical approximation.

For outgoing modes in the noncommutative model, 
the wave equation~\eqref{eom_2p} implies the dispersion relation
\be 
\label{disp_rel}
\omega = \frac{p}{2} \left[1 - \frac{a}{r} \, \Theta\biggl(v - \frac{\ell^2 |p|}{2} \biggr)\right]
=
\begin{dcases}
\frac{p}{2} \left(1 - \frac{a}{r}\right)
,
\quad 
& v > \ell^2 |p| / 2
\\
\frac{p}{2}
\, ,
\quad 
& v < \ell^2 |p| / 2
\end{dcases}
\, ,
\ee 
where $\omega$ is the eigenvalue of the operator $i\partial_v$. 
Compared with the commutative case, 
the only effect of noncommutativity on the wave equation 
is to shift the advanced-time location 
of the collapsing null shell by $\Delta v = \ell^2 |p|/2$.

Let us consider an outgoing Hawking particle outside the shell 
and trace its phase-space trajectory $(r(v), p(v))$ 
backward in the advanced time $v$. 
The dispersion relation~\eqref{disp_rel} defines the point-particle Hamiltonian $\omega=\omega(r,p)$, 
from which the semiclassical trajectory follows Hamilton's equations:
\begin{align}
\label{Hamilton_eq}
\frac{dr}{dv} = \frac{\del \omega}{\del p} = \frac{1}{2}\left(1 - \frac{a}{r}\right) \, ,
\qquad
\frac{dp}{dv} = - \frac{\del \omega}{\del r} = - \frac{ap}{2r^2} \, .
\end{align}

Our main interest is in the $v$-dependence of the particle momentum $p$. 
In the black-hole region outside the shell, 
it evolves as
\be 
\label{p(v)}
p(v) =
2 \omega \, 
\frac{1 + W[ e^{(v \, - \, u_0) / 2 a}]}
{W[ e^{(v \, - \, u_0) / 2 a}]}
\ee 
according to eq.~\eqref{Hamilton_eq},
where $W(z)$ denotes the Lambert-$W$ function, 
and $u_0$ labels the retarded time of the outgoing massless particle.
For a distant observer located at fixed radial distance $r_0\simeq (v_0-u_0)/2\gg a$, 
the particle momentum~\eqref{p(v)} at detection (namely at $v=v_0$) 
is approximately $p (v_0) \simeq 2 \omega$, in agreement with 
the dispersion relation~\eqref{disp_rel} in the asymptotically flat region.
On the other hand, when the particle is traced back in time
toward the near-horizon region where $v \ll u_0$, 
the expression~\eqref{p(v)} reduces to
\be 
p (v)
\simeq 
2 \omega \, 
e^{(u_0 \, - \, v) / 2a}
\, .
\ee

In the commutative case, where the collapsing shell is located at $v=0$, 
a Hawking particle detected at late retarded time $u_0\gg a$ 
therefore has momentum
\be 
\label{ps0}
p (v = 0)
\simeq 
2 \omega \, 
e^{u_0 / 2a}
\ee 
on the shell, which is matched to its momentum $p(v<0)$ 
in the flat region inside the shell. 
This exponential dependence of the interior momentum 
on the asymptotic retarded time $u_0$ 
is the origin of Hawking radiation at the Hawking temperature $T_{\mathrm{H}}=1/4\pi a$.

In the noncommutative case, however, 
the shell is shifted to $v = \ell^2 |p|/2$, 
and the particle momentum $p_s \equiv p(v = \ell^2 |p_s| / 2)$ 
at the shifted shell is now determined near the horizon 
by the algebraic equation 
\be 
p_s \simeq 
2 \omega \, 
\exp \! \left( \frac{u_0 - \ell^2 p_s / 2}{2a} \right) 
.
\ee 
Solving this equation gives
\be 
\label{ps}
p_s
\simeq 
\frac{4a}{\ell^2} \, 
W\biggl(\frac{\ell^2 \omega}{2a} \, e^{u_0 / 2a} \biggr)
\, .
\ee 
In the commutative limit $\ell\to 0$, 
this expression reproduces the standard exponential blueshift~\eqref{ps0}. 
In the opposite regime where 
\be 
\label{u0_linear}
\omega \, e^{u_0 / 2a}
\gg 
\frac{2a}{\ell^2}
\, ,
\ee 
one instead finds
\be 
\label{ps_scr}
p_s
\simeq 
\frac{4a}{\ell^2} \, 
\log\biggl(\frac{\ell^2 \omega}{2a} \, e^{u_0 / 2a} \biggr)
\simeq 
\frac{2 u_0}{\ell^2}
\, .
\ee 
Thus, the momentum inside the shell grows only \emph{linearly} 
with the asymptotic retarded time $u_0$ for large $u_0$. 
For a typical Hawking particle with frequency $\omega\sim \mathcal{O}(1/a)$, 
the condition~\eqref{u0_linear} corresponds to 
\be 
u_0 \gg u_{\mathrm{scr}} 
\equiv 
2a 
\log \biggl( \frac{a^2}{\ell^2} \biggr) \, ,
\label{u0>uscr}
\ee 
where $u_{\mathrm{scr}}$ coincides with the scrambling time
$\mathcal{O}\!\left(a\log(a^2/\ell^2)\right)$. 
As a result, Hawking radiation is expected to be suppressed 
beyond the scrambling time $u \gg u_{\mathrm{scr}}$, 
which is indeed what we will later demonstrate in section~\ref{sec:HR_NC}.

For an outgoing particle, 
the momentum inside the shell determines 
the effective shell position in advanced time through $v_s = \ell^2 p_s/2$.
Since the Eddington retarded and advanced times $(u, v)$
are related to the radial coordinate by $v - u = 2r + 2a \log(r/a - 1)$
(\emph{cf.} eq.~\eqref{u(v,r)}),
the corresponding effective radial position $r_s$ of the shell satisfies
\be 
r_s (u_0) - a  
\; \simeq \; 
\frac{\ell^2 a \omega}{eu_0} 
\ll \mathcal{O}\left(\frac{\ell^2}{a}\right)
\ee 
for a typical Hawking particle with $\omega\sim \mathcal{O}(1/a)$ 
at retarded time $u_0\gg u_{\mathrm{scr}}$. 
Hence, even though a late-time Hawking particle 
with $u_0\gg u_{\mathrm{scr}}$ perceives a large shift
$\Delta v \simeq u_0 \gg a$ in the shell location along the $v$-direction,
the effective shell position nevertheless remains 
well within the near-horizon region.\footnote{
For comparison, in the commutative case 
the radial position of the collapsing null shell as a function of $u_0$ is
$r_s (u_0) - a \simeq a \, e^{-u_0/2a-1}$ near the horizon.
}

\section{Hawking Radiation in Noncommutative Spacetime}
\label{sec:HR_NC}

The derivation of Hawking radiation in the noncommutative Vaidya background 
closely parallels that in the commutative case reviewed in section~\ref{sec:HR0}. 
The noncommutative wave equation~\eqref{eom_2p} can be written as
\be 
\begin{aligned}
r \, (2 \del_v + ip) \, \tilde{\phi}(v, p) 
= ia \, 
\Theta\biggl(v - \frac{\ell^2 |p|}{2}\biggr) \, p \, \tilde{\phi} (v, p) \, ,
\label{eom_2p_r}
\end{aligned}
\ee 
where $r$ is an operator acting on functions of $p$.

Outside the shifted shell $v > \ell^2 |p|/2$,
the single-frequency solution $\tilde{\phi}_{\omega}(v,p)$ 
of the deformed wave equation~\eqref{eom_2p_r} 
coincides with that in the commutative case,
which is given by eq.~\eqref{phi0_om} in the near-horizon region.
Following the same steps as in section~\ref{sec:sol0}, 
we consider the field configuration 
\be
\tilde{\phi} (v > \ell^2 |p|/2 \, , p)
=
\widetilde{\Psi}_{\omega_c , \, u_c} (v, p)
\equiv 
\int_{-\infty}^{\infty} 
d \omega \, 
\widetilde{\psi}_{\omega_c}(\omega) \, 
e^{i \omega u_c} \, 
\tilde{\phi}_{\omega}(v, p)
\ee 
for $v > \ell^2 |p|/2$, where $\widetilde{\Psi}_{\omega_c , \, u_c}(v, p)$ represents the wave packet of a Hawking particle with central frequency $\omega_c \sim \mathcal{O}(1/a) > 0$, localized around the retarded time $u = u_c$.

As in the commutative case, 
we determine the Bogoliubov decomposition of the wave-packet annihilation operator $\mathfrak{b}_{\Psi} \equiv \langle \Psi_{\omega_c, \, u_c} \, , \phi \rangle$ 
in terms of the Minkowski creation and annihilation operators 
$(\mathfrak{a}_p \, ,\mathfrak{a}_p^\dagger)$ inside the shell.
The number expectation value of Hawking particles 
associated with the wave packet $\Psi_{\omega_c,u_c}$ 
is then given by the Klein-Gordon norm~\eqref{inner} of the negative-$p$ components 
of the wave packet inside the shell:
\be 
\label{bbvev}
\ev{\mathfrak{b}_{\Psi}^{\dagger} \mathfrak{b}_{\Psi}}{0} 
=
\int_0^{\infty} dp \, 
\left( 2 p \right) 
\abs\Big{
\widetilde{\Psi}_{\omega_c , \, u_c} 
\bigl( 
\ell^2 \abs{p} / 2 \,
- p
\bigr)}^2
\, .
\ee
The only difference from the commutative expression~\eqref{bbvev0} 
is that the matching is now performed at 
the momentum-dependent shell interface $v=\ell^2 |p|/2$.

Evaluating the single-frequency solution~\eqref{phi0_om} 
at the shifted shell location yields 
\be 
\tilde{\phi}_{\omega}(v = \ell^2 |p|/2 \, , p)
\equiv 
\mathcal{N}_{\omega} \, 
e^{-i \ell^2 \omega |p|/2} \, 
e^{- iap} 
\left[ 
a \left( p - i \varepsilon \right) 
\right]^{-1 \, - \, 2 i a \omega} 
\, .
\ee 
Accordingly, the negative-momentum components of the wave packet
entering the integrand of~\eqref{bbvev} has the explicit form
\be 
\widetilde{\Psi}_{\omega_c , \, u_c} 
\bigl( 
\ell^2 \abs{p} / 2 
\, , 
- p
\bigr)
=
\frac{e^{i a \abs{p}}}{- a \abs{p}} \, 
\int_{-\infty}^{\infty} d \omega \,
\widetilde{\psi}_{\omega_c} (\omega) \,
\mathcal{N}_{\omega} \, 
e^{-2 \pi a \omega} \, 
\exp 
\left\{ 
i \omega 
\left[ 
u_0 (\abs{p}; u_c) - \frac{\ell^2 |p|}{2}
\right]
\right\} 
,
\ee 
where we have used the function $u_0 (p; u_c) \equiv u_c - 2 a \log(ap)$ 
introduced in eq.~\eqref{p_to_u(p)}.
Applying the narrow-band approximation described in eq.~\eqref{packet_approx} suitable for a sharply peaked frequency profile $\widetilde{\psi}_{\omega_c} (\omega)$, the slowly varying factor $\mathcal{N}_{\omega} \, e^{-2 \pi a \, \omega}$ may be treated as approximately constant and pulled out from the integral as $\mathcal{N}_{\omega_c} \, e^{-2 \pi a \, \omega_c}$.
This allows us to further simplify 
\begin{align}
\abs\Big{
\widetilde{\Psi}_{\omega_c , \, u_c} 
\bigl( 
\ell^2 \abs{p} / 2 
\, , 
- p
\bigr)}^2
\nn
&\simeq 
\frac{a}{e^{4 \pi a \, \omega_c} - 1} \, 
\frac{1}{p^2} \, 
\bigg\lvert 
\int_{-\infty}^{\infty} 
\frac{d \omega}{\sqrt{2 \pi}} \,
\widetilde{\psi}_{\omega_c} (\omega) \, 
e^{i \omega \, [ u_0 (\abs{p}; \, u_c) \, - \, \ell^2 \abs{p} / 2 ]}
\bigg\rvert^2
\\
&= 
\frac{a}{e^{4 \pi a \, \omega_c} - 1} \, 
\frac{1}{p^2} \, 
\bigl| \psi_{\omega_c} \bigl( u(p; u_c) \bigr)
\bigr|^2
\qquad 
(p > 0)
\, ,
\label{AbsPsi}
\end{align}
where $\psi_{\omega_c} (u)$ denotes the inverse Fourier transform of the frequency profile $\widetilde{\psi}_{\omega_c} (\omega)$ (\emph{cf.} eq.~\eqref{psi(u)}).
Here we have also defined the function
\be 
\label{u(p)}
u(p; u_c)
\equiv 
u_0 (p; u_c) - \frac{\ell^2}{2} \, p
=
u_c
-
2a \log (a p) 
- 
\frac{\ell^2}{2} \, p
\qquad 
(p > 0)
\, ,
\ee 
which serves as the noncommutative analogue 
of the low-energy expression $u_0 (p; u_c)$~\eqref{p_to_u(p)}. 
The additional term $- \ell^2 p / 2$ arises from 
the momentum-dependent shift $v = \ell^2 \abs{p} / 2$ 
of the effective advanced-time location of the shell.

Substituting eq.~\eqref{AbsPsi} into eq.~\eqref{bbvev}, 
we arrive at 
\begin{align}
\ev{\mathfrak{b}_{\Psi}^{\dagger} \mathfrak{b}_{\Psi}}{0} 
&\simeq
\frac{2 a}{e^{4 \pi a \, \omega_c} - 1}
\int_0^{\infty} 
\frac{dp}{p} \, 
\bigl| 
\psi_{\omega_c} \bigl( u(p; u_c) \bigr)
\bigr|^2
\nn \\
&=
\frac{1}{e^{4 \pi a \, \omega_c} - 1}
\int_{-\infty}^{\infty}
du \, 
\frac{1}{1 + \ell^2 p(u; u_c) / 4a} \, 
\bigl| 
\psi_{\omega_c} (u)
\bigr|^2
\, ,
\label{bb_final}
\end{align}
where the function 
\be 
p(u; u_c)
\equiv 
\frac{4a}{\ell^2} \, 
W \biggl( \frac{\ell^2}{4a^2} \, e^{(u_c \, - \, u) / 2a} \biggr)
\label{puuc}
\ee 
is obtained by inverting eq.~\eqref{u(p)}. 
This is reminiscent of eq.~\eqref{ps} in the semiclassical approximation.
For Hawking wave packets whose central retarded time lies in the regime  
$u_c \gg u + 2a \log(a^2 / \ell^2)$,  
the asymptotic behavior of the Lambert-$W$ function implies 
$p(u; u_c) \simeq 2 \, (u_c - u) / \ell^2$, 
which again closely resembles eq.~\eqref{ps_scr} in the semiclassical analysis.

Compared to the result~\eqref{bb_final0} in the commutative case, 
the noncommutative correction to Hawking radiation 
is encoded entirely in the nontrivial measure factor
$\left[ 1 + \ell^2 p(u; u_c) / 4a \right]^{-1}$ appearing in the integrand of eq.~\eqref{bb_final}.
This factor indicates that the magnitude of Hawking radiation 
acquires an explicit dependence on the retarded time parameter $u_c$ 
of the detected Hawking wave packet, 
and the Hawking flux therefore becomes \emph{time-dependent} 
from the viewpoint of a distant observer.

The profile $\psi_{\omega_c} (u)$ in the integrand of eq.~\eqref{bb_final} is, by definition, localized around $u = 0$ (\emph{cf.} eq.~\eqref{psi(u)}).
On the other hand, one observes from eq.~\eqref{puuc} that 
\be 
p(u = 0; u_c)
\simeq 
\frac{2}{\ell^2} \, u_c
\qquad 
\text{for}
\quad 
u_c \gg u_{\mathrm{scr}} 
\equiv 
2a 
\log \biggl( \frac{a^2}{\ell^2} \biggr)
\, .
\ee 
Hence, near $u=0$, the measure factor behaves as
\be 
\frac{1}{1 + \ell^2 p(0; u_c) / 4a}
\simeq 
\frac{2a}{u_c}
\ll 
1
\qquad 
\text{for}
\quad 
u_c \gg u_{\mathrm{scr}} 
\, .
\ee 
At sufficiently late times, the measure decays linearly with $u_c$, indicating that the number of Hawking particles emitted well beyond the scrambling time is strongly suppressed:
\be 
\label{bb_suppressed}
\ev{\mathfrak{b}_{\Psi}^{\dagger} \mathfrak{b}_{\Psi}}{0} 
\sim
\frac{1}{e^{4\pi a \, \omega_c} - 1} \,
\frac{2a}{u_c}
\to 0
\qquad 
\text{for}
\quad 
u_c \gg u_{\mathrm{scr}} 
\, .
\ee 

Recall that, in the commutative case, the Hawking radiation of a single massless scalar field  (\emph{cf.} eq.~\eqref{bb_final0}) leads to the decay rate~\cite{Page:1976df}
\be
\label{decay_rate0}
\frac{da(u)}{du} 
\simeq 
- \frac{1}{1920 \pi} \, 
\frac{\ell_p^2}{a^2(u)} \, ,
\ee
where $\ell_p$ denotes the Planck length. 
Although $\ell_p$ need not coincide with the noncommutativity length scale $\ell$, 
we assume for simplicity that they are of the same order: 
$\mathcal{O}(\ell_p)\sim \mathcal{O}(\ell)$. 
A conventional black hole therefore evaporates 
over a timescale of order the Page time: 
$\Delta u \simeq 640 \pi \, a_0^3/\ell_p^2$, 
where $a_0$ is the initial Schwarzschild radius.

In the noncommutative model considered here, 
the decay rate is suppressed relative to the standard result~\eqref{decay_rate0}
by a factor $2a/u$ after the scrambling time 
(\emph{cf.} eq.~\eqref{bb_suppressed}), \emph{i.e.}, 
\be 
\label{da/du_NC}
\frac{da(u)}{du}
\simeq
- \frac{\alpha}{2 \pi} \, 
\frac{\ell_p^2}{u \, a(u)}
\qquad 
\text{for}
\quad 
u \gg u_{\mathrm{scr}} 
=
\mathcal{O} \bigl( a_0 \log(a_0^2 / \ell^2) \bigr)
\, ,
\ee
where $\alpha$ is a constant that incorporates the greybody factor 
as well as the weighted number of massless degrees of freedom.\footnote{
Ignoring the greybody factor, we have $\alpha = N/480$ for $N$ scalars~\cite{Page:1976df}.
}
Note that the radiation emitted before the scrambling time 
removes only a negligible fraction of the total mass:
\be
\label{Deltaaovera}
\frac{\Delta a(u_{\mathrm{scr}})}{a_0} 
\sim 
\mathcal{O}\!\left( \frac{\ell_p^2}{a_0^2} \, \log(a_0^2 / \ell^2) \right) \ll 1 \, .
\ee
Therefore, on timescales much longer than the scrambling time, 
eq.~\eqref{da/du_NC} implies that $a(u)$ is approximately given by
\be 
a^2(u) 
\simeq
a^2(u_{\mathrm{scr}}) 
- 
\frac{\alpha}{\pi} \, 
\ell_p^2 \, 
\log \biggl(\frac{u}{u_{\mathrm{scr}}} \biggr)
\qquad \mbox{for} \;\;  u \gg u_{\mathrm{scr}} \, .
\ee

Consequently, spacetime noncommutativity predicts a dramatically prolonged  
evaporation time
\be 
\label{lifetime}
u_{\mathrm{evap}}
\simeq
u_{\mathrm{scr}} \, 
\exp \biggl[
\frac{\pi}{\alpha} \, \frac{a^2(u_{\mathrm{scr}})}{\ell_p^2} 
\biggr] \, ,
\ee
which is exponentially longer than the Page time $\mathcal{O}(a_0^3 / \ell_p^2)$.
Since the derivation above retains only the leading terms 
in the expansion with respect to $\ell^2/a^2$, 
subleading corrections of order 
$\mathcal{O} \bigl( (\ell^2 / a^2) \, \log(a^2/\ell^2) \bigr)$
in the exponent of eq.~\eqref{lifetime} 
modify the prefactor $u_{\mathrm{scr}}$ 
by a factor of $(\ell^2/a^2)^m$ with some power $m$. 
For this reason, it is more meaningful to 
suppress the prefactor 
and express the evaporation time parametrically as
\be 
\label{lifetime-1}
u_{\mathrm{evap}}
\sim
\exp \biggl(\frac{\pi \, a_0^2}{\alpha \, \ell_p^2} \biggr) 
\sim \exp(\frac{\mathcal{S}_{\mathrm{BH}}}{\alpha}) \, ,
\ee
where 
\be
\mathcal{S}_{\mathrm{BH}} 
\equiv 
\frac{A}{4 \, \ell_p^2}
\ee
is the Bekenstein-Hawking entropy of the black hole,
with $A = 4\pi a_0^2$ being the initial horizon area.

The evaporation timescale $\sim e^{\mathcal{S}_{\mathrm{BH}} /\alpha}$ 
has the same exponential form as the Poincar\'e recurrence time 
$\sim e^{\mathcal{S}_{\mathrm{BH}}}$. 
In our physical universe, 
both timescales are so enormous that they are effectively irrelevant 
for macroscopic systems. 
In practice, one expects classical or quantum instabilities 
to occur on much shorter timescales. 
In this respect, the present scenario is not drastically different 
from UV models in which Hawking radiation terminates around the scrambling time~\cite{Barman:2017vqx, Ho:2022gpg, Chau:2023zxb, Ho:2023tdq, Ho:2024tby}, 
since those models are likewise expected to be subject to instabilities 
long before the recurrence time is reached.

\newpage 
\section{Conclusion and Outlook}
\label{sec:conclude}

In this work, we have employed a noncommutative field-theory model for the radiation field to investigate potential UV effects on Hawking radiation 
from a Schwarzschild black hole formed by the collapse of a null thin shell. 
Our analysis reveals that spacetime noncommutativity 
deforms the gravitational interaction at the UV scale $\ell^{-1}$, 
resulting in a substantial suppression of Hawking radiation. 
In particular, for distant observers, 
the radiation intensity decays as $1/u$ in retarded time $u$
after the scrambling time $\mathcal{O}(a\log(a^2/\ell^2))$ 
(see eq.~\eqref{bb_suppressed}). 
This in turn implies an exponentially prolonged black-hole lifetime
$\sim e^{\mathcal{S}_{\mathrm{BH}}/\alpha}$, 
where $\mathcal{S}_{\mathrm{BH}}=\pi a^2/\ell^2$ denotes the black-hole entropy 
and $\alpha$ is an $\mathcal{O}(1)$ constant. 
This timescale far exceeds the conventional evaporation time $\mathcal{O}(a^3/\ell^2)$.

Although the evaporation time remains finite, 
it is essentially of the order of the \emph{Poincar\'e recurrence time}. 
We therefore argue that the information paradox may be alleviated, 
since the black hole now has a vastly extended period 
over which information can be transferred from the interior to the exterior 
through other mechanisms, 
such as classical decay or quantum tunneling, 
which are expected to appear within the recurrence time. 
It should be emphasized that the information paradox 
concerns Hawking radiation specifically, 
which arises from the reinterpretation of the vacuum 
inside the collapsing matter shell by distant observers.

The spacetime noncommutativity in our model 
does not significantly modify low-energy effective physics 
for either freely falling or distant observers, 
and is therefore not excluded by any known experiment. 
Moreover, generalizing the setup to a smooth, non-constant noncommutativity 
is not expected to qualitatively change the conclusions of this work, 
as long as a finite noncommutativity remains near the horizon, 
since each late-time Hawking wave packet probes only
a tiny neighborhood of the horizon. 
Likewise, small deformations of the Vaidya geometry 
(\emph{e.g.}, replacing the null thin shell by a thick shell 
collapsing at subluminal speed) 
should not alter the qualitative picture, 
provided that a horizon still forms 
and the collapsing velocity remains of order unity near the horizon.

Similar suppression effects on Hawking radiation due to UV physics 
have previously been found in a model that implements 
the generalized uncertainty principle in the radiation field~\cite{Chau:2023zxb},
as well as in studies~\cite{Ho:2023tdq, Chang:2024scn} based on the covariant spacetime uncertainty relation $\Delta u\,\Delta v \gtrsim \ell^2$.
Those models predict an evaporation that stops shortly after the scrambling time, 
with only negligible loss of black-hole mass~\cite{Chau:2023zxb, Ho:2023tdq, Ho:2024tby}. 
By contrast, although the noncommutative model considered here 
also incorporates a spacetime uncertainty relation $\Delta v\,\Delta r \gtrsim \ell^2$, 
the suppression is milder, 
and the black hole is expected to evaporate completely within a finite time.
To our knowledge, this is the first example of a UV model 
featuring spacetime uncertainty that nevertheless yields a finite evaporation time.
Despite the differences in mechanism and detailed evaporation dynamics among these UV modifications, they all point to a common qualitative conclusion: 
string-inspired uncertainty relations~\cite{Yoneya:1987gb, Amati:1988tn, Yoneya:1989ai, Konishi:1989wk, Guida:1990st, Yoneya:1997gs, Yoneya:2000bt} that exhibit UV-IR connection lead to an $\mathcal{O}(1)$ suppression of Hawking radiation around the scrambling time.

In the standard picture of Hawking radiation, 
a particle emitted at later retarded time 
originates from a region outside the collapsing shell 
that lies closer to the horizon, 
and therefore undergoes an exponentially larger gravitational blueshift. 
This exponential blueshift is essential 
for maintaining a constant radiation flux. 
In the UV model studied here, 
the noncommutative interaction between the radiation field 
and the Vaidya background effectively shifts the thin shell at $v=0$ 
to a momentum-dependent location $v_{s}(p) = \ell^2 |p|/2$,
with the shift proportional to the momentum $p$ of the outgoing mode. 
Since Hawking quanta detected at later times 
correspond to larger momenta when traced back toward the horizon, 
the effective displacement of the shell also becomes larger at later times. 
As a result, the blueshift relation is substantially softened in the late-time regime, leading to a large modification of Hawking radiation after the scrambling time. 
Our results therefore highlight the importance of incorporating the collapsing matter in the analysis of Hawking radiation.

Despite the infinitely many $v$-derivatives 
appearing in the star product used to implement spacetime noncommutativity, 
the deformed field equation considered in this work 
differs from its commutative counterpart in the Vaidya spacetime 
only through the momentum-dependent shift $\Delta v=\ell^2 |p|/2$ 
in the location of the collapsing shell. 
This simplified model admits a Hermitian Hamiltonian (see \emph{e.g.}, eq.~\eqref{H}), 
making the unitarity of the wave equation manifest.

The time dependence of the intensity of Hawking radiation is not only of theoretical interest, but also of phenomenological relevance. 
In particular, it plays an important role in the observational prospects of primordial black holes and their viability as dark matter candidates. 
Recent studies~\cite{Alexandre:2024nuo, Dvali:2024hsb, Dvali:2025ktz} have shown that, under the proposal that Hawking radiation is suppressed around the Page time $\mathcal{O}(a^3 / \ell^2)$ due to the ``memory burden’’ effect~\cite{Dvali:2018xpy, Dvali:2020wft}, a new mass window opens for light primordial black holes.
In view of our finding that the black-hole lifetime is exponentially long $\sim e^{\mathcal{O}(1)\,a^2/\ell^2}$, even Planck-scale black holes could in principle contribute to the dark matter abundance today.\footnote{
For illustration, a black hole with mass $10$ times the Planck mass would have an evaporation timescale~\eqref{lifetime} of order 
$\ell_p \, e^{\mathcal{S}_{\mathrm{BH}} / \alpha} \sim 10^{506}$ years for $\alpha=1$.
}
A detailed study of these phenomenological implications is left for future work.

\section*{Acknowledgements}

We thank Katsuki Aoki, Sinya Aoki, Jil Le Bois, Johanna Borissova, Chong-Sun Chu, Norihiro Iizuka, Akihiro Ishibashi, Satoshi Iso, Hyun Jeong, Hikaru Kawai, Alessandro Manta, Masamichi Miyaji, Cheng-Tsung Wang, and Yuki Yokokura for valuable discussions. 
P.M.H. is supported in part by the Ministry of Science and Technology, 
R.O.C. (NSTC 112-2112-M-002-024-MY3, NSTC 113-2112-M-002-040-MY2), 
and by National Taiwan University. 
W.H.S. received support from the Ministry of Science and Technology, R.O.C. 
(NSTC 112-2112-M-002-024-MY3, NSTC 113-2112-M-002-040-MY2), and National Taiwan University during the early stages of this work, 
and was supported by the Special Postdoctoral Researcher (SPDR) Program at RIKEN 
during the latter stages of its completion. 
T.Y. is supported in part by JSPS KAKENHI Grant No.~JP22H05115 and 25K17390.

\small

\bibliographystyle{myJHEP}
\bibliography{bibliography}

\end{document}